# Enabling High-Power, Broadband THz Generation with 800-nm Pump Wavelength

Zachary B. Zaccardi[†], Isaac C. Tangen[†], Gabriel A. Valdivia-Berroeta, Charles B. Bahr, Karissa C. Kenney, Claire Rader, Matthew J. Lutz, Brittan P. Hunter, David J. Michaelis*, and Jeremy A. Johnson*.

Department of Chemistry and Biochemistry, Brigham Young University, Provo, UT, 84602, USA.
*email: dmichaelis@chem.byu.edu, jjohnson@chem.byu.edu

**Abstract**

The organic terahertz (THz) generation crystal BNA has recently gained traction as a valuable source to produce broadband THz pulses. Even when pumped with 800-nm light, thin BNA crystals can produce relatively high electric fields with frequency components out to 5 THz. However, the THz output when pumped with 800-nm light is limited by the damage threshold of the organic crystal. Here we report that the damage threshold of BNA can be significantly improved by physically bonding BNA to a high-thermal conductivity sapphire window. When pumped with 800-nm light from an amplified Ti:sapphire laser system, our bonded BNA (BNA-sapphire) generates 2.5× higher electric field strengths compared to bare BNA crystals. We characterize the average damage threshold for bare BNA and BNA-sapphire, measure peak-to-peak electric field strengths and THz waveforms, and determine the nonlinear transmission in BNA. Pumping BNA-sapphire with 800-nm light results in peak-to-peak electric fields exceeding 1 MV/cm, with strong broadband frequency components from 0.5-5 THz. Our BNA-sapphire THz source is a promising alternative to tilted pulse front LiNbO$_3$ THz sources, which will enable many research groups without optical parametric amplifiers to perform high-field, broadband THz spectroscopy.

## 1. Introduction

High-field terahertz (THz) science is enabling fascinating new studies of condensed matter systems, including powerful recent applications of two-dimensional THz spectroscopy [1-7]. The generation of high-field THz radiation is most commonly achieved using high-power, ultrafast laser pulses derived from amplified Ti:sapphire laser systems. To date, nonlinear THz studies that require strong peak electric fields (>100 kV/cm) have either been carried out a) by directly pumping inorganic LiNbO$_3$ using a tilted-pulse front configuration [8,9], or b) by using organic THz generation crystals like DAST [10], DSTMS [11], OH1 [12], and EHPSI-4NBS [13], with broader generation bandwidths. LiNbO$_3$ may be pumped directly with the 800-nm output of a Ti:sapphire laser producing THz with relatively low frequencies (<3 THz). Organic THz generators can produce broader bandwidths (1-6 THz), however they are generally more efficiently pumped at longer wavelengths, often necessitating an optical parametric amplifier (OPA) to produce these longer pump wavelengths. With pump pulse energies of a few millijoules, peak-to-peak electric field strengths exceeding 1 MV/cm can be achieved in either THz generation scheme, enabling a host of powerful measurements [1-4,6,7,14-18].

For many ultrafast laser research groups without an optical parametric amplifier, high-field THz science has been limited to lower frequencies (< 3 THz) accessible with LiNbO$_3$. However, if an OPA is available, organic crystals have higher THz generation efficiencies that make up for power losses in the OPA generation of 1200-1600 nm light. Furthermore, the good phase-matching in organic crystals enables a simple collinear THz generation geometry. This also allows for easily changing the polarization of generated THz by simply rotating the THz generation crystal along with the polarization of the pump light. In short, while LiNbO$_3$ currently best utilizes the Ti:sapphire output of 800-nm, organic crystals can offer advantages of broader generation bandwidths, easy polarization control, and simple THz generation geometries when an OPA is available. Therefore, finding an organic nonlinear optical crystal that offers these advantages without the need of an OPA to generate pump light would be a boon to many THz research groups.

We have recently demonstrated that the thin crystals of organic BNA can generate broad bandwidth THz pulses when pumped with the 800-nm light output of an amplified Ti:sapphire laser [19]. However, the low melting point of BNA (103º C) leads to a relatively low laser induced damage threshold (LIDT) of 4 mJ/cm$^2$ when pumped with ~100 fs ultrafast pulses of 800-nm light at a 500 Hz repetition rate. In addtion, the damage threshold drops to 2 mJ/cm$^2$ when BNA is pumped at 1 kHz repetition rate, suggesting that the damage mechanism involves the crystal heating to its relatively low melting point and melting. Another paper reports a LIDT for BNA of 6 mJ/cm$^2$ when using 50 fs 800-nm pulses at 100 Hz repetition rate [20]. These damage thresholds prohibit many Ti:sapphire laser systems from utilizing their full power to pump BNA crystals for THz generation.

In this report, we show that bonding BNA to high-thermal conductivity sapphire plates significantly increases the damage threshold and greatly boosts the THz output. The increased heat dissipation raises the damage threshold of the BNA almost threefold, which also increases the attainable electric field by a factor of ~2.4. The use of BNA-sapphire therefore can thus enable high-field, broadband THz spectroscopy in any research lab with an amplified, ultrafast Ti:sapphire laser system, and produce broader THz spectra than is possible with LiNbO$_3$.

## 2. Methods

Bare BNA crystals and BNA-sapphire structures were produced by Terahertz Innovations, LLC. In the BNA-sapphire structures, BNA is physically bonded to a sapphire plate such that there are no glues, no air gaps, or other material between the crystal and the sapphire, which can reduce heat dissipation and increase Fresnel losses and lower performance [21]. BNA crystals with thicknesses less than ~300 μm were classified as "thin" (most were approximately 200 μm thick), and BNA crystals with thicknesses greater than ~300 μm were classified as "thick" (most were approximately 400-500 μm thick). Thin and thick crystals were mounted bare or fused to 0.5-mm thick, 1-inch diameter sapphire plates to test the impact of the sapphire layer on the THz generation output of BNA. We classified the four types of samples tested as thin bare BNA, thick bare BNA, thin BNA-sapphire, and thick BNA-sapphire. Larger BNA crystals (14 mm × 25 mm) were irradiated with fluences up to the damage threshold multiple times so that damage spots did not overlap. The LIDT of each sample type was measured at least five times.

To perform the THz generation experiments, two optical rectification setups were utilized. In each, ultrafast laser pulses were generated by a Ti:sapphire system with a central wavelength at 800 nm. The pulse duration was ~100 fs with a 500 Hz repetition rate (mechanically chopped from 1 kHz). In the first setup, the pump beam was reduced to a 2.6 mm 1/e$^2$ radius. 800-nm pulses were directed to the BNA or BNA-sapphire structures, a Teflon filter removed the remaining pump light, and the generated THz naturally diverged to nearly fill and was focused by a 2-inch diameter off-axis parabolic mirror. The focused THz (~500 μm radius) waveform was detected with 100-fs 800-nm probe pulses in 100 μm (110) GaP layer bonded to a 1 mm (100) GaP crystal to minimize the effect of signal echoes. In the second setup, the pump beam was larger with a 4.5 mm 1/e$^2$ radius, and a series of 3 off-axis parabolic mirrors were used to focus the THz beam to approximately 300 μm radius with the same EO-detection scheme.

Typical THz waveforms with accompanying spectra measured in the first setup are displayed in **Fig. 1**. In **Figs. 1a** and **1b**, the lighter lines were produced with bare BNA crystals, and the darker lines were generated with BNA bonded to sapphire (thin crystals for **Fig. 1a** and thicker crystals for **Fig. 1b**). Thicker crystals exhibit a generation spectrum similar to LiNbO$_3$ (up to about 3 THz), and thinner crystals contain significant frequency content out to 5 THz.

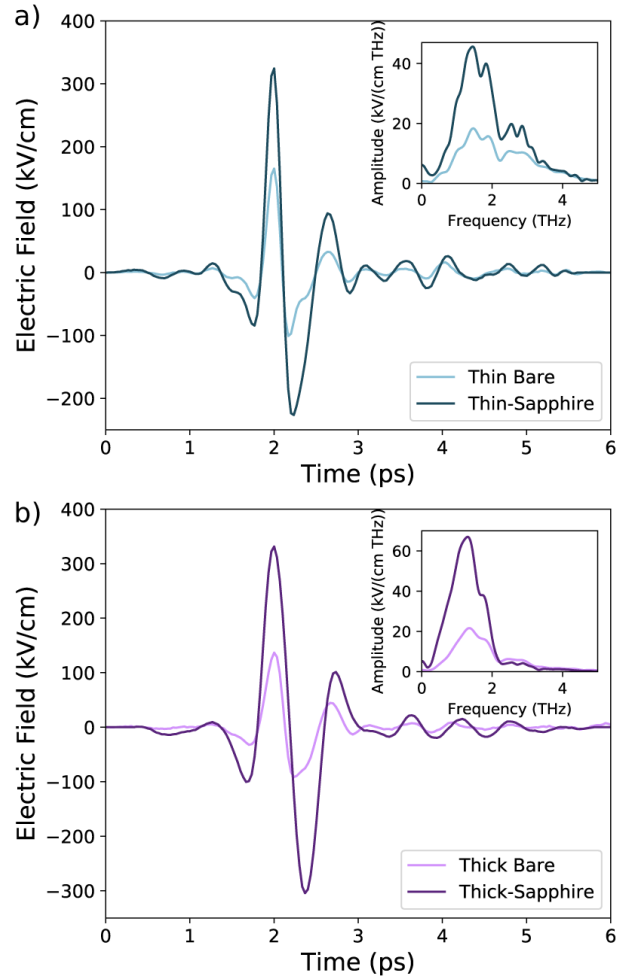

**Figure 1** Generated THz electric fields for bare crystals and BNA-sapphire structures. a) thin bare BNA crystal (light blue) and thin BNA-sapphire structure (navy blue). b) thick bare BNA crystal (light purple) and thick BNA-sapphire structure (purple).

## 3. Results and discussion

**Table 1** shows representative photos and diagrams of the tested architectures, with the corresponding LIDT values and maximum generated electric field results (before damage). To test for the damage threshold, the sample was exposed to the laser for at least 15 seconds at each fluence. After increasing the fluence, a drop in the output THz electric field was an indication that the sample had been damaged. We report average damage threshold values of 4.84 ± 0.95 and 4.03 ± 0.17 mJ/cm$^2$ for thin and thick bare crystals, respectively. At these fluence values, bare BNA crystals heat and melt, leaving a hole in the BNA crystals that results in a significant decrease of THz signal (see **Fig. 2a**). The obtained LIDT results are in agreement with previously reported damage threshold values of 4 and 6 mJ/cm$^2$ [19,20]. The BNA-sapphire structures, however, maintain their integrity up to an average fluence of 13 mJ/cm$^2$. Even at 16 mJ/cm$^2$, BNA-sapphire structures show damage as a browning of the crystal with no melted holes, as illustrated in

**Fig. 2b.** After damage and lowering the fluence below the damage threshold, the BNA-sapphire structures still produced strong THz, but the electric-field output was reduced by ~10%.

**Table 1** Representative photos and diagrams of thin and thick bare BNA and fused to sapphire crystals. Average and maximum laser-induced damage threshold (LIDT) values were determined for a ~2.6 mm pump radius. Maximum electric field values were obtained for crystal spots irradiated at high fluences.

| Sample | Thin Bare | Thin-Sapphire | Thick Bare | Thick-Sapphire |
|---|---|---|---|---|
| | 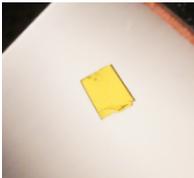 | 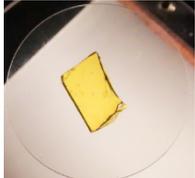 | 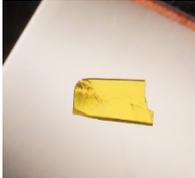 | 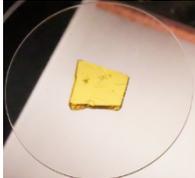 |
| | 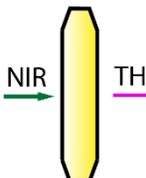 | 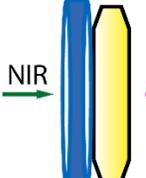 | 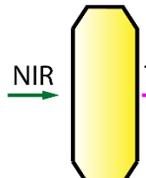 | 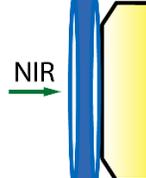 |
| Av. LIDT (mJ/cm$^2$) | 4.84 ± 0.95 | 13.2 ± 2.2 | 4.03 ± 0.17 | 11.6 ± 2.4 |
| Max. LIDT (mJ/cm$^2$) | 6.2 | 16.0 | 4.2 | 14.9 |
| Electric field (kV/cm) | 238 | 551 | 266 | 636 |

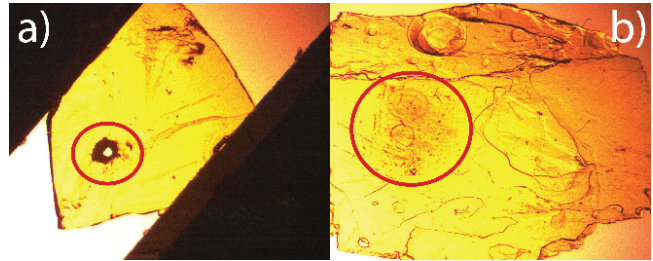

**Figure 2** Damaged BNA crystals, a) Bare BNA, and b) BNA-sapphire structure. Damaged spots are marked with red circles.

The marked increase in the LIDT of the BNA-sapphire structures corresponds on average to 2.8 times higher stability compared to bare BNA crystals for both thin and thick samples. To demonstrate the reproducibility of a) our THz generation measurements and b) the BNA-sapphire structures, **Table 1** reports the average damage threshold for bare BNA and BNA-sapphire samples. The bare thin crystals show a larger standard error compared with the thick samples and similar uncertainties are observed for crystal-sapphire measurements.

To demonstrate the impact of the featured BNA-sapphire structures on the generated THz fields, we show in **Fig. 3** the evolution of the peak electric fields with increasing pump fluences. Importantly, the bare crystals reach their damage threshold at around a fluence of 4 mJ/cm$^2$, when the crystals begin to melt and the THz signal significantly decreases. For the fused structures, the magnitude of the THz electric field continues to increase up to about a fluence of 12 mJ/cm$^2$, on average. While there is not a linear relationship between pump fluence and output THz electric field, an increase in pump fluence nonetheless results in an increase in generated electric field until the BNA sample is damaged. As seen from **Fig. 3**, an increase of about ~3 times in the fluence with the crystal-sapphire ensemble results in around 2.4-fold improvement in the generated THz fields.

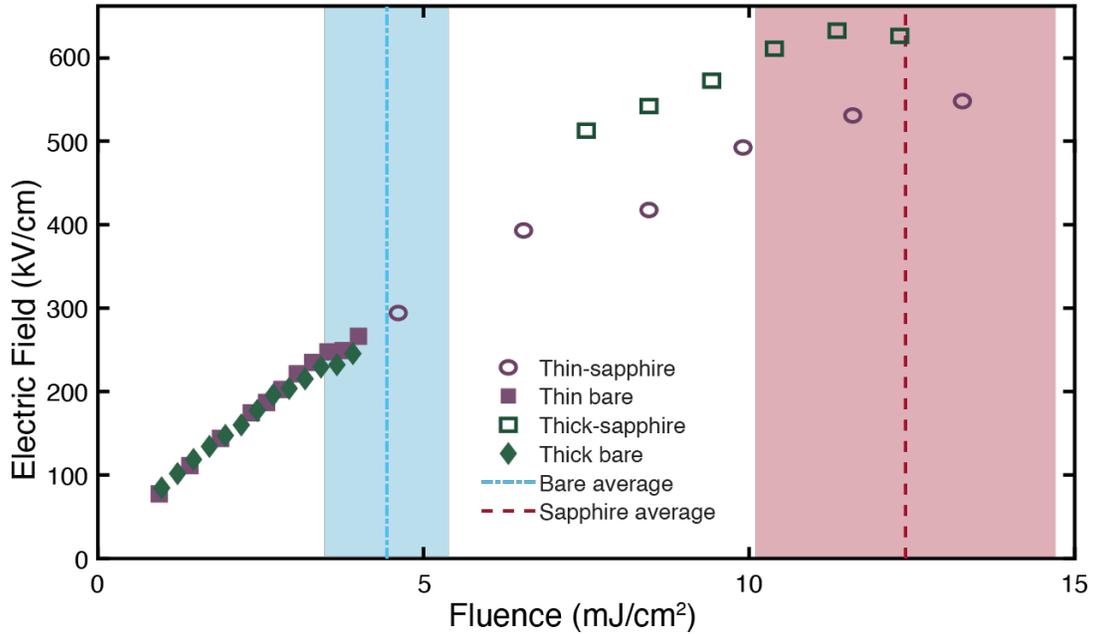

**Figure 3** Electric field values at different fluences for representative crystal spots. Bare and crystal-sapphire samples are represented with filled and open shapes, respectively. The dashed and dot-dashed lines represent the LIDT, while the shaded portions represent standard deviations of the LIDT.

**Figure 1** above shows THz waveforms that were recorded for using thin and thick crystal structures at fluences just below the laser-induced damage threshold. As seen in **Fig. 1a** and **Fig 3**, a thin crystal-sapphire structure pumped at 13.2 mJ/cm$^2$ generates a peak to peak THz field of 551 kV/cm, and as observed in **Fig. 1b** and **Fig. 3**, the thick BNA-sapphire generates 636 kV/cm. These peak fields more than double the maximum THz fields of 238 and 266 kV/cm obtained for bare thin and thick crystals, respectively (also in **Figs. 1a** and **1b**). The THz generation spectral data for bare crystals and crystal-sapphire structures are shown in the insets of **Fig, 1,** showing no large change in the frequency content of the generation spectra. For all crystals, the amplification caused by the increased pump fluence reaches a maximum at 1.4 THz, where phase-matching is optimal [19]. The broadband generation for the thin BNA-sapphire architecture is similar to our previous study with thin BNA crystals pumped at 800 nm and extends up to 5 THz [19], while the thick BNA-sapphire crystals produce very strong fields below 2 THz, with smaller amounts of light produced up to 3 THz.

As a further test of the capabilities enabled by the sapphire structures, we changed the setup to pump the BNA-sapphire structures with the natural laser beam radius of 4.5 mm, with full laser power of 4.2 mJ (corresponding to a fluence of 6.6 mJ/cm$^2$). In addition, better focusing was achieved by using three parabolic mirrors instead of one [8]. As shown in **Fig. 4**, we achieved a peak to peak electric field of over 1 MV/cm with the thick BNA sample. At these peak electric fields, the response in GaP begins to saturate, and our measurements place a lower bound on the electric field strength. The inset to **Fig. 4** shows the normalized spectra with frequency components extending out to 5 THz. These field strengths now rival those produced with tilted-pulse front THz generation in LiNbO$_3$, but with a simple collinear pump and a broader THz bandwidth.

Furthermore, the broadband THz pulses generated with BNA-sapphire now rival the strength of other organic THz generators pumped at longer wavelengths. With similar pump fluences, BNA is not as efficient at THz generation as DAST or OH1, but with the ability to use more power directly from a Ti:sapphire laser, the overall BNA THz output is now similar to more efficient THz generators. In **Fig. 4**, we also show THz pulses generated using DAST or OH1 pumped at 1450 nm. These THz traces were measured using exactly the same 3-parabolic mirror experimental setup for all crystals. DAST and OH1 were pumped by the maximum power output of our OPA with 0.84 mJ per pulse, while BNA is pumped with 4.2 mJ of 800-nm light. The lower THz generation efficiency of BNA-sapphire is similar to the power conversion losses in the OPA needed to efficiently pump DAST and OH1.

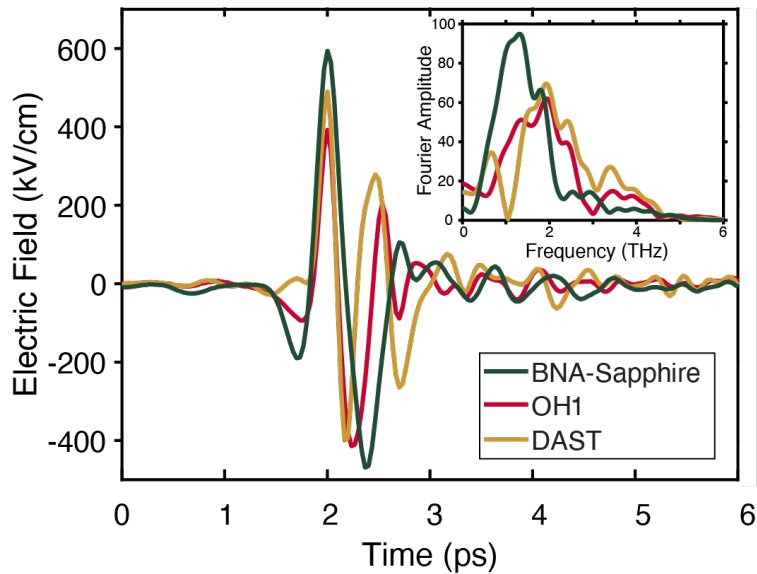

**Figure 4.** THz waveforms collected with BNA-sapphire pumped with near full laser power, compared to THz generated with full output power from the OPA used to pump DAST or OH1 crystals. Inset: Corresponding spectra.

To further investigate this increase in damage threshold, we analyzed the transmission of 800-nm light through bare BNA and BNA-sapphire as a function of fluence. A large 10 mm × 20 mm, ~650 μm thick BNA crystal was fused to a sapphire plate, half on and half off so we could directly compare transmission through the bare crystal to transmission through the BNA-sapphire. **Figure 5** shows the transmission that nonlinearly drops as the fluence is increased for both the bare BNA and BNA-sapphire. The nonlinear transmission is explained well with a multiphoton absorption fit (dashed line in **Fig. 5**).

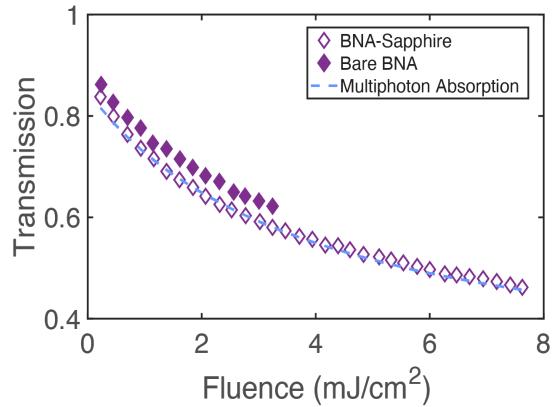

**Figure 5** Transmittance of the 800-nm pump through bare BNA and BNA-sapphire as a function of pump fluence. The dashed line shows a fit to a multiphoton absorption model, including two- and three-photon absorption.

As the structure absorbs more and more pump light, the BNA heats up more and more. Using the nonlinear absorption (adjusted for reflective losses using a refractive index of 1.76 for sapphire, 1.836 for BNA, and 1 for air), and estimating the thermal conductivity of BNA as 0.18 W/mK (which is similar to many organic crystals), we can approximate the steady-state temperature rise of the BNA and compare this to the temperature rise of the BNA-sapphire structure (see equations 6 and 8 from Ref. [22]). For bare BNA, the calculations predict a temperature increase of over 80 K, which would bring BNA above the melting point at the center of the pump beam, resulting in a melted hole in the BNA (see **Fig. 2a** above). We can also calculate the temperature change of BNA fused to sapphire. When fused to high-thermal conductivity sapphire, we estimate BNA only heats up ~5-10 K, even at the highest fluences reached. Imaging the BNA crystals with a thermal imaging camera confirms this temperature rise of only a few degrees for the BNA-sapphire and ~70 degree rise for the bare BNA. If we consider that the steady-state temperatures of the BNA-sapphire structures are well below the melting point, we conclude that a different damage mechanism is occurring when the BNA-sapphire browns (see **Fig. 2b** above). Fortunately, this browning-damage is not catastrophic, and the BNA can still be used, albeit with slightly less THz output.

## 4. Conclusions

In summary, we have shown the utility of fusing BNA crystals to high-thermal conductivity sapphire windows to improve thermal stability, significantly increase the damage threshold, and enable large increases in THz output via pumping with Ti:sapphire laser sources. The 3-fold increase in damage threshold for BNA-sapphire ensembles compared to bare crystals is due to improved heat dissipation from the BNA into the sapphire layer. This increased damage threshold enables significantly increased THz generation. With larger spot sizes and pump pulse energies, broadband THz pulses with peak to peak electric fields of over 1 MV/cm are attainable. The BNA-sapphire generated THz fields and broad bandwidths are comparable in magnitude to those produced by the organic crystals DAST and OH1, which are optimally pumped using the longer wavelength output of an optical parametric amplifier. These BNA-sapphire structures will enable any lab with an amplified Ti:sapphire laser system to now perform high-field THz spectroscopy with broader bandwidth THz pulses than are possible with tilted-pulse front THz generation with $LiNbO_3$.

†Contributed equally to this work.

**Acknowledgements.** We thank the Department of Chemistry and Biochemistry at Brigham Young University for funding.

**Disclosures.** D. J. Michaelis and J. A. Johnson are co-founders of Terahertz Innovations, LLC.